# Bottom-up Engineering of Diamond Nanostructures

Igor Aharonovich[1,2,*], Jonathan C. Lee[1], Andrew P. Magyar[1,3], David O. Bracher[1] and Evelyn L. Hu[1]

1. School of Engineering and Applied Sciences, Harvard University, Cambridge, MA, 02138 , USA
2. School of Physics and Advanced Materials, University of Technology Sydney, Broadway, NSW, 2007, Australia
3. Center for Nanoscale Systems, Harvard University, Cambridge, MA, 02138, United States
* igor@seas.harvard.edu

**Abstract**
*Engineering nanostructures from the bottom up enables the creation of carefully engineered complex structures that are not accessible via top down fabrication techniques, in particular, complex periodic structures for applications in photonics and sensing. In this work, we propose and demonstrate a bottom up approach that can be adopted and utilized to controllably build diamond nanostructures. A realization of periodic structures and optical wave-guiding is achieved by growing nanoscale single crystal diamond through a defined pattern.*

Controlled engineering of materials into ordered periodic arrays at the nanoscale enables manipulation of the interaction between light and matter[1-3]. In particular, realization of nanophotonic devices such as photonic crystal cavities enables investigation of quantum effects including light-matter interaction or enhancement of spontaneous emission rates, as well as practical applications in sensing, optical nonlinearity and lasing[4, 5]. To achieve optimal performance, the optical cavities have to be physically isolated to confine light at the nanoscale. Therefore, their fabrication requires thin, wavelength-sized suspended membranes and typically demands lithographic procedures involving multiple etching steps. Indeed, the majority of devices for photonics, sensing and nano-electronics have been engineered predominantly from materials such as silicon, silica and GaAs.

Recently, diamond has emerged as a promising platform for a variety of nanophotonic and sensing applications owing to its exceptional physical and mechanical properties, chemical inertness and its ability to host ultra-bright single photon emitters[6-11]. However, diamond is not amenable to routine, scalable nano-engineering and processing, mainly due to lack of native oxide and its chemical stability. Recent reports on generating devices by ion implantation[12], bonding approach[13] or angle etching[14] showed excellent geometrical structures, however, the final devices still require long reactive ion etching (RIE) steps that can hinder the quality of the material. Furthermore, some device geometries are not readily accessible via a directional dry etch.

In nature, bottom up self-assembly yields complex photonic structures[15-18]. Such structures have been mimicked synthetically with materials including polymers and silica spheres[2, 15]. Bottom up approaches have also been utilized for silicon[19] and GaAs[20], however, expending this approach to diamond is challenging due to the complex required growth conditions. Unlike other semiconductors, diamond cannot be grown by MBE, nor can it be grown epitaxially on a sacrificial non-diamond substrate. Furthermore, diamond growth occurs isotropically in three dimensions and, consequently, growth of high aspect ratio structures (e.g. nanowires) is challenging. Finally, diamond growth occurs at elevated temperatures under high plasma densities, inhibiting the use of polymers as mask during growth. Herein, we realize periodic nanostructures out of single crystal diamond with excellent optical and structural properties fabricated by growth using microwave-enhanced plasma chemical vapor deposition (MPCVD).

The bottom up structure is grown through a silica template on a single crystal diamond, type IIA, [100] oriented (Element Six). 300 nm of $SiO_2$ is deposited on the top surface to serve as a hard mask. E-beam lithography is used to pattern a periodic structure, which is transferred to the $SiO_2$ mask. RIE in fluorine environment is used to etch the pattern into the $SiO_2$ and expose the diamond surface. Figure 1a shows a scanning electron microscope (SEM) image of the mask. The growth of the patterned diamond is carried out in a MPCVD system under the following conditions: pressure 60 torr, microwave plasma 950 W, $CH_4$:$H_2$ 400:4 standard cubic centimeter per minute. The patterned structure aims to limit the diamond growth laterally and provide faster growth in the vertical direction. Figure 1b shows an SEM image of the diamond sample with the patterned mask after the growth. The growth replicates the periodicity of the pattern with excellent precision. Each patterned aperture yields only one diamond crystal. By employing a high hydrogen to methane ratio and relatively high pressure (~60 torr), secondary nucleation is suppressed and the diamond nucleates occurs only on the exposed diamond surface. After growth, the $SiO_2$ mask is removed by immersing the sample in hydrofluoric acid. Figure 1c shows an example of another pattern with square periodicity and smaller features after the HF treatment. Inhomogeneous nucleation is suppressed and the grown crystals are faceted, having uniform size and morphology while forming a perfect long-range array of nanoscale crystals. Inset is a higher magnification image of several diamond nanostructures.

One of the major advantages of the bottom-up approach, particularly for diamond, is engineering more complex geometries that are not possible with top down etching. Figure 1d shows an example of a single crystal diamond pyramid grown through an aperture on top of a single crystal diamond. The top of the pyramid is suspended above the diamond surface, forming an isolated photonic constituent. This geometry is achieved in two stages: first the lateral growth is confined by the aperture and diamond grows vertically. Second, the diamond reaches the height of the mask and grows both laterally and vertically, forming the pyramid geometry. Similar strategy can be used to grow diamond bridges and other undercut geometries, and indicates the possibility of growing more complex geometries such as inverse opal structures.

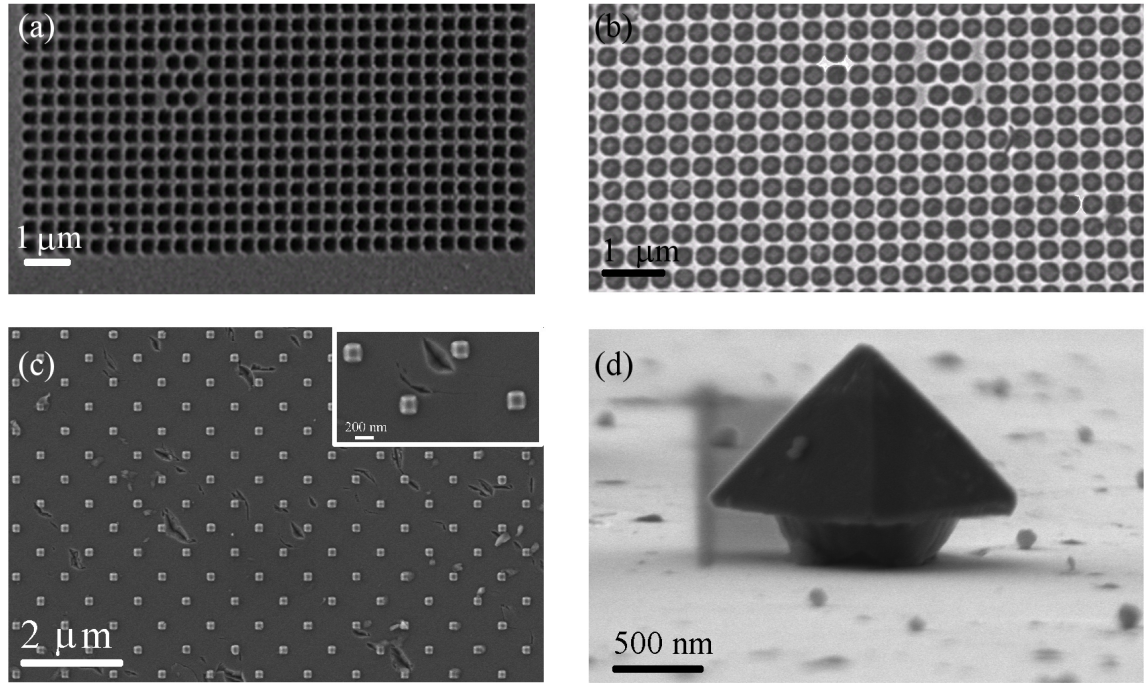

*Figure 1. (a) SiO$_2$ hard mask is patterned on the top [100] diamond surface. (b) A single crystal nanodiamond array is grown through the patterned SiO$_2$ hard mask. (c) Example of a different pattern after SiO$_2$ removal – exposing perfectly aligned periodic array of diamond nanostructures. (d) An individual diamond nanocrystal is formed from overgrowing the SiO$_2$ mask.*

While the diamond nanostructures above are fabricated on the surface of bulk diamond, for many photonic applications, an isolated configuration, traditionally a thin suspended membrane, waveguides or microrings are resting on a low index material – such as SiO$_2$. To demonstrate the suitability of this bottom up technique for practical photonic applications, waveguide structures and hexagonal microdisks[21] are patterned. To provide the necessary optical isolation, thin diamond membranes are utilized as templates.

Diamond membranes, 1.7 μm thick, are produced by ion implantation and lift-off, as reported previously[22]. A thin layer of SiO$_2$ (200 nm) is deposited on the membrane using PECVD and the desired pattern is defined by e-beam lithography. The membrane is then introduced into an MPCVD reactor for the "bottom up" growth of the waveguides and microdisks. After the growth, the hybrid device is flipped over and the original membrane template is removed by RIE in oxygen plasma, leaving behind only the grown photonic structures. Fig 2a illustrates this procedure. Fig 2b shows an optical photograph of the stand-alone diamond devices (bright features), resting on a SiO$_2$ substrate. Fig 2c shows SEM images of hexagonal microdisks array and a diamond waveguide.

The optical properties of the grown resonators were characterized using a 532 nm excitation laser at room temperature. Fig 2c shows a PL spectrum recorded from one of the microdisks, showing a narrowband emission centered at 738 nm, which is attributed to the SiV defect. The SiV defect attracts considerable attention recently, due to its narrow emission line, short excited state lifetime and exceptional brightness[10, 23]. Hence, the bottom up fabrication of resonators containing this defect is a promising route towards practical realization of integrated diamond photonics. Remarkably, no other color centers are grown into the optical resonator, endowing with a high signal to noise ratio and an efficient filtering. Inset of fig 2c is a Raman spectrum recorded from the same microdisk. A sharp line at 1332 cm$^{-1}$ (2.5 cm$^{-1}$ Full Width at Half Maximum), the first order diamond Raman line, is clearly seen, indicating the high quality of the grown resonator. Note that the final etch step to remove the original diamond membrane does not degrade the grown structures. Furthermore, any damage is localized on the top surface of the structure rather than on the edges, minimizing overlap between potential RIE damage and the modes of the device

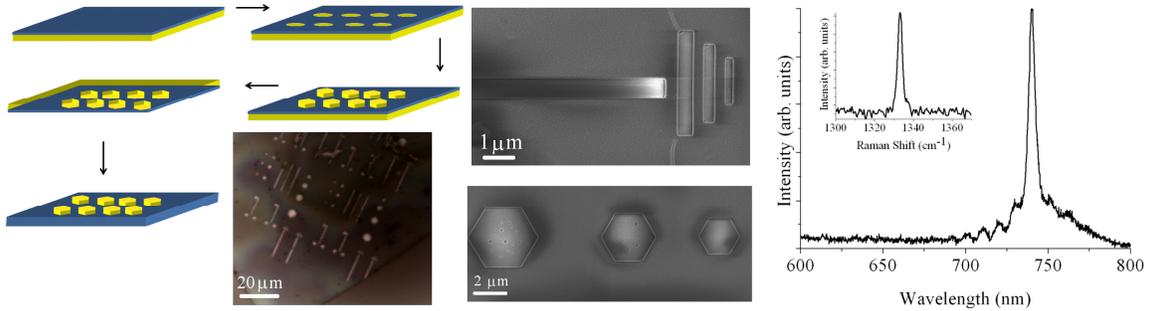

*Figure 2. (a) Schematic illustration of the bottom up approach to generate optical resonators out of diamond membranes. A thin layer of SiO$_2$ (blue) is deposited on a diamond membrane (yellow). The pattern is defined by e-beam lithography and the membrane is transferred for the regrowth. Then, the structure is flipped and the original diamond membrane is removed by RIE, leaving behind pristine single crystal diamond resonators. (b) Optical image of the bottom-up cavities and waveguides. Bright features are the diamond resonators, while the surrounding is the SiO$_2$ substrate. (c) SEM images of an array of hexagonal resonators and a diamond waveguide (d) PL spectrum recorded from one of the resonators using 532 nm excitation. Inset, Raman spectrum showing a sharp line at 1332 cm$^{-1}$, characteristic to a first order diamond Raman line.*

The fabricated waveguides are further characterized through a transmission measurement. Figure 3a shows an optical image of the diamond waveguide. To carry out the transmission measurement, an excitation and the collection spots were separated, as indicated by the black and red arrow, respectively. Fig 3b shows the resulted PL from this measurement. The transmitted SiV signal through the waveguide is clearly visible, indicating a proper working waveguide. Some losses are expected since the diamond is positioned on top of a SiO$_2$ substrate. Nevertheless, this is an unprecedented demonstration of bottom up engineering of photonic constituents from diamond.

The bottom up growth of diamond nanostructures is a powerful tool to generate homogeneous diamond nanostructures with excellent optical performance. There are several advantages of the bottom up approach for fabricating structures from diamond. First, this approach enables generation of photonic constituents that did not undergo any ion damage or etching damage during the process, and therefore is expected to yield the best quality material. Second, employing the bottom up approach, nanodiamonds with narrow size distribution and homogeneous geometries can be engineered – useful for photonic applications such as pillar photonic crystal cavities. Finally, the bottom up technique has the potential to controllably engineer various color centers within the nanodiamonds – a powerful tool for many biomarking applications and multicolor imaging. The generation of color centers does not require ion implantation, and hence the whole device is not amenable to ion damage.

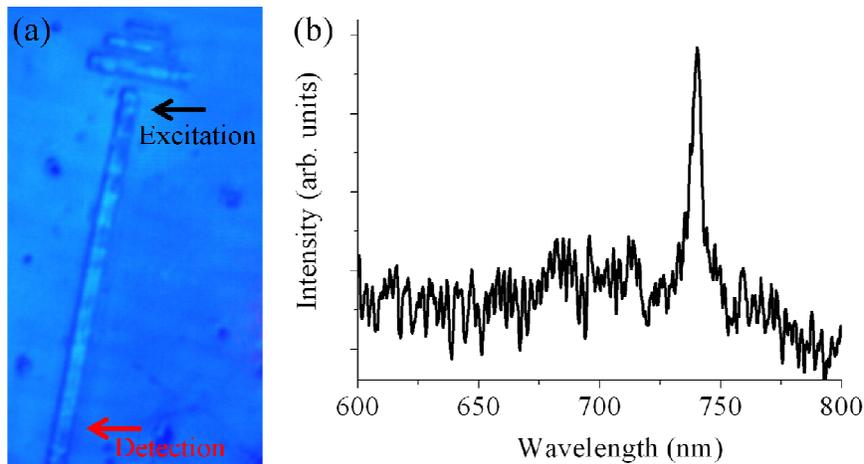

*Figure 3. (a) Optical image of the bottom up engineered diamond waveguide. (b) Transmission spectrum recorded from this waveguide with excitation and detection positions as indicated by the black and red arrows, respectively.*

In summary we demonstrate the bottom up growth of oriented, single crystal diamond nanostructures. Realization of a periodic array of single crystal nanodiamonds on top of a bulk crystal diamond is a highly sought after feature for applications in sensing and anti-reflection coatings. We further envision that pillar photonic crystal cavities grown from diamond may be extremely promising for applications in bio-sensing. The bio-compatibility of diamond concurrent with surface pillars hosting optically active defects, can be utilized for inter cellular electrical signals[24], as the fluid can easily flow in between the pillars.

Thin diamond membranes were utilized to grow periodic structures – including hexagonal optical resonators and waveguides. This work opens new avenues for exploring novel fabrication techniques with diamond, to take advantage of the ease of growth of one-dimensional structures, and enables more complex 3D structures that are not accessible via traditional top down approaches.

The authors acknowledge T.L. Liu and Thomas Babinec for helpful discussions and Prof David Clarke for the access to the Raman facilities. The financial support of the DARPA under the Quantum Entanglement Science and Technology (QuEST) Program. This work was performed in part at the Center for Nanoscale Systems (CNS), a member of the National Nanotechnology Infrastructure Network (NNIN), which is supported by the National Science Foundation under NSF Award No. ECS-0335765. CNS is part of the Faculty of Arts and Sciences at Harvard University.